\documentclass[conference]{IEEEtran}
\usepackage{cite}

\ifCLASSINFOpdf
\usepackage[pdftex]{graphicx}
\else
\fi
\usepackage{amsmath}
\usepackage{amsfonts} 
\usepackage{amsthm}
\usepackage{amssymb}
\usepackage[boxruled]{algorithm2e}
\usepackage{amsthm} 
\usepackage{mathtools} 
\usepackage{textcomp} 

\usepackage{BOONDOX-ds}

\usepackage{bbm}

\newcommand{\pfun}{\mathop{\hbox{$\to$\kern-7pt\raise.9pt\hbox{\scalebox{1}[.55]{$|$}}\kern4pt} }}

\usepackage{array}

\begin{document}

\title{75,000,000,000 Streaming Inserts/Second Using \\ Hierarchical Hypersparse GraphBLAS Matrices}

\author{\IEEEauthorblockN{Jeremy Kepner$^{1,2,3}$, Tim Davis$^{4}$, Chansup Byun$^1$, \\ William Arcand$^1$, David  Bestor$^1$, William Bergeron$^1$, 
 Vijay Gadepally$^{1,2}$, Matthew Hubbell$^1$, \\ Michael Houle$^1$, Michael Jones$^1$, Anna Klein$^1$, Peter Michaleas$^1$, Lauren
    Milechin$^5$, \\ Julie Mullen$^1$, Andrew Prout$^1$, Antonio Rosa$^1$,   Siddharth Samsi$^1$, Charles
Yee$^1$, Albert Reuther$^1$
\\
\IEEEauthorblockA{$^1$MIT Lincoln Laboratory Supercomputing Center, $^2$MIT Computer Science \& AI Laboratory, \\ $^3$MIT Mathematics Department, $^4$Texas A\&M, $^5$MIT Department of Earth, Atmospheric and Planetary Sciences}}}
\maketitle

\begin{abstract}
The SuiteSparse GraphBLAS C-library implements high performance hypersparse matrices with bindings to a variety of languages (Python, Julia, and Matlab/Octave).  GraphBLAS provides a lightweight in-memory database implementation of hypersparse matrices that are ideal for analyzing many types of network data, while providing rigorous mathematical guarantees, such as linearity.  Streaming updates of hypersparse matrices put enormous pressure on the memory hierarchy.  This work benchmarks an implementation of hierarchical hypersparse matrices that reduces memory pressure and dramatically increases the update rate into a hypersparse matrices.  The parameters of hierarchical hypersparse matrices rely on controlling the number of entries in each level in the hierarchy before an update is cascaded.  The  parameters are easily tunable to achieve optimal performance for a variety of applications.   Hierarchical hypersparse matrices achieve over 1,000,000 updates per second in a single instance.  Scaling to 31,000 instances of hierarchical hypersparse matrices arrays on 1,100 server nodes on the MIT SuperCloud achieved a sustained update rate of 75,000,000,000 updates per second.  This capability allows the MIT SuperCloud to analyze extremely large streaming network data sets.  
\end{abstract}

%
\IEEEpeerreviewmaketitle

\section{Introduction}
\let\thefootnote\relax\footnotetext{This material is based upon work supported by the Assistant Secretary of Defense for Research and Engineering under Air Force Contract No. FA8702-15-D-0001 and National Science Foundation CCF-1533644. Any opinions, findings, conclusions or recommendations expressed in this material are those of the author(s) and do not necessarily reflect the views of the Assistant Secretary of Defense for Research and Engineering or the National Science Foundation.}

The global Internet is expected to exceed 100 terabytes per second (TB/s) by the year 2022 creating significant performance challenges for the monitoring necessary to improve, maintain, and protect the Internet, particularly with the rising social influence of  adversarial botnets encompassing a significant fraction of Internet traffic \cite{cisco2018cisco,allcott2017social,BadBotReport}.  Origin-destination traffic matrix databases are fundamental network analysis tool for a wide range of networks, enabling the observation of temporal fluctuations of network supernodes, computing background models, and inferring the presence of unobserved traffic \cite{soule2004identify, zhang2005estimating, bharti2010inferring, tune2013internet, 8916263}.  Rapidly constructing these traffic matrix databases is a significant productivity, scalability, representation, and performance challenge \cite{castellana2017high, busato2018hornet,8547563,8547570,8547514,8547572,samsi2017static,kao2017streaming}.

  Our team has developed a high-productivity scalable platform---the MIT SuperCloud---for providing scientists and engineers the tools they need to analyze large-scale dynamic data \cite{kepner2012dynamic,gadepally2018hyperscaling,8547629}.  The MIT SuperCloud provides interactive analysis capabilities  accessible from high level programming environments (Python, Julia, Matlab/Octave) that scale to thousands of processing nodes.   Traffic matrices can be manipulated on the MIT SuperCloud using distributed databases (SciDB and Apache Accumulo), D4M associative arrays \cite{kepner2012dynamic,kepnerjananthan}, and now the SuiteSparse GraphBLAS hypersparse matrix library \cite{7761646,Davis:2019:ASG:3375544.3322125,8547538}.
  
  For IP network traffic data, the IP address space requires a hypersparse matrix (\#entries $<<$ \#rows and \#columns) that is either $2^{32}{\times}2^{32}$ for IPv4 or $2^{64}{\times}2^{64}$ for IPv6. Our prior work represented traffic matrices using D4M associative arrays using sorted lists of strings to describe the row and column labels of an underlying standard sparse matrix \cite{8916508}.  D4M associative arrays provide maximum flexibility to represent the row and columns with arbitrary strings and are extremely useful during the feature discovery stage of algorithm development.  For IP traffic matrices, the row and column labels can be constrained to integers allowing additional performance to be achieved using a hypersparse matrix library.  In either case, the memory hierarchy presents a significant performance bottleneck as doing lots of updates to slow memory is prohibitive.  This work benchmarks an implementation of hierarchical hypersparse matrices that reduces memory pressure and dramatically increases the update rate into a hypersparse matrices.

\begin{figure*}[]
\centering
\includegraphics[width=6in]{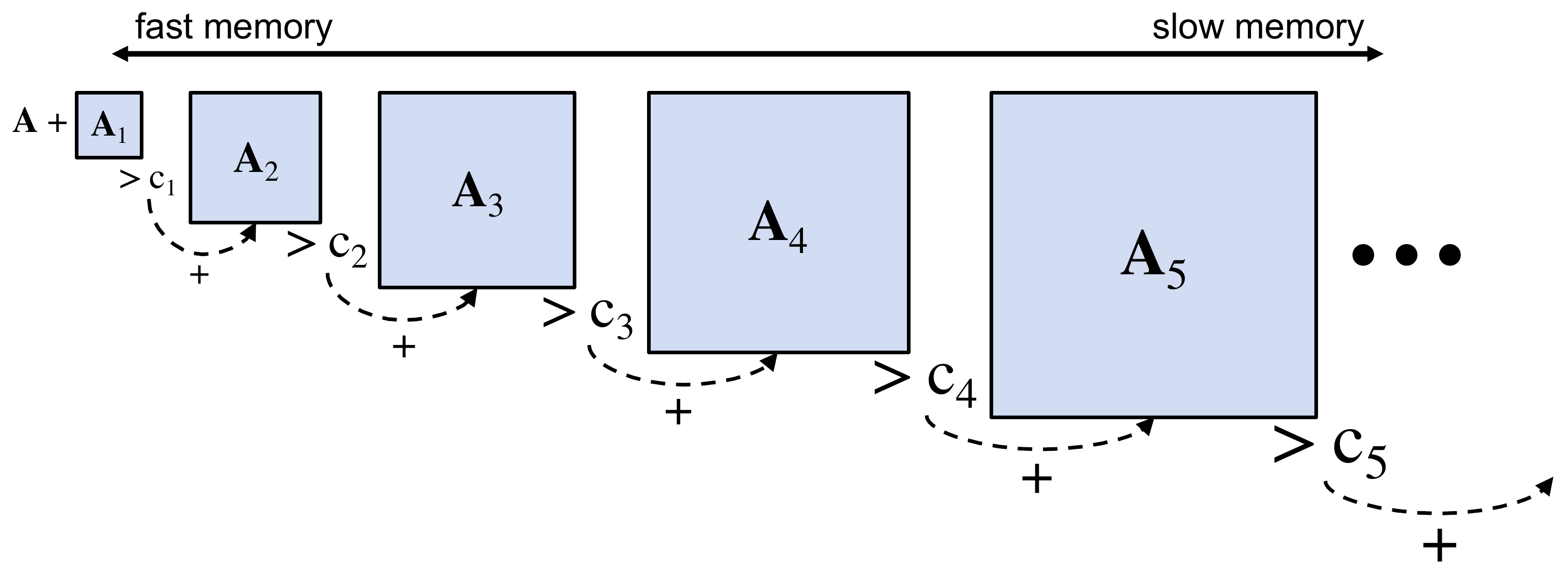}
\caption{Hierarchical hypersparse matrices store increasing numbers of nonzero entries in each layer (adapted from \cite{8547629}).  If layer ${\bf A}_i$ surpasses the nonzero threshold $c_i$ it is added to ${\bf A}_{i+1}$ and cleared.  Hierarchical hypersparse matrices ensure that the majority of updates are performed in fast memory.}
\label{fig:HierarchicalArrays}
\end{figure*}

\section{Hierarchical Hypersparse Matrices}

The SuiteSparse GraphBLAS library is an OpenMP accelerated C implementation of the GraphBLAS.org sparse matrix standard.  Python, Julia, and Matlab/Octave bindings allow the performance benefits of the SuiteSparse GraphBLAS C library to be realized in these highly productive programming environments.  Streaming updates to a large hypersparse matrix can be be accelerated with a hierarchical implementation optimized to the memory hierarchy (see Fig.~\ref{fig:HierarchicalArrays}).   Rapid updates are performed on the smallest hypersparse matrices in the fastest memory.  The strong mathematical properties of the GraphBLAS allow a hierarchical implementation of hypersparse matrices to be implemented via simple addition. All creation and organization of hypersparse row and column indices are handled naturally by the GraphBLAS mathematics.  If the number of nonzero (nnz) entries exceeds the threshold $c_i$, then ${\bf A}_i$ is added to ${\bf A}_{i+1}$ and ${\bf A}_i$ is cleared.  The overall usage is as follows
\begin{itemize}
\item Initialize $N$-level hierarchical hypersparse matrix with cuts $c_i$
\item Update by adding data ${\bf A}$ to lowest layer
$$
  {\bf A}_1 = {\bf A}_1 + {\bf A}
$$
\item If ${\rm nnz}({\bf A}_1) > c_1$, then
$$
  {\bf A}_2 = {\bf A}_2 + {\bf A}_1
$$
and reset ${\bf A}_1$ to an empty hypersparse matrix of appropriate dimensions.
\end{itemize}

\noindent The above steps are repeated until ${\rm nnz}({\bf A}_i) \leq c_i$ or $i = N$.  To complete all pending updates for analysis, all the layers are added together
$$
   {\bf A} = \sum_{i=1}^{N} {\bf A}_i
$$
Hierarchical hypersparse matrices dramatically reduce the number of updates to slow memory.  Upon query, all layers in the hierarchy are summed into the hypersparse matrix.  The cut values $c_i$ can be selected so as to optimize the performance with respect to particular applications.  The majority of the complex updating is performed by using the existing GraphBLAS addition operation.

\section{Scalability Results}

\begin{figure}[]
\centering
\includegraphics[width=\columnwidth]{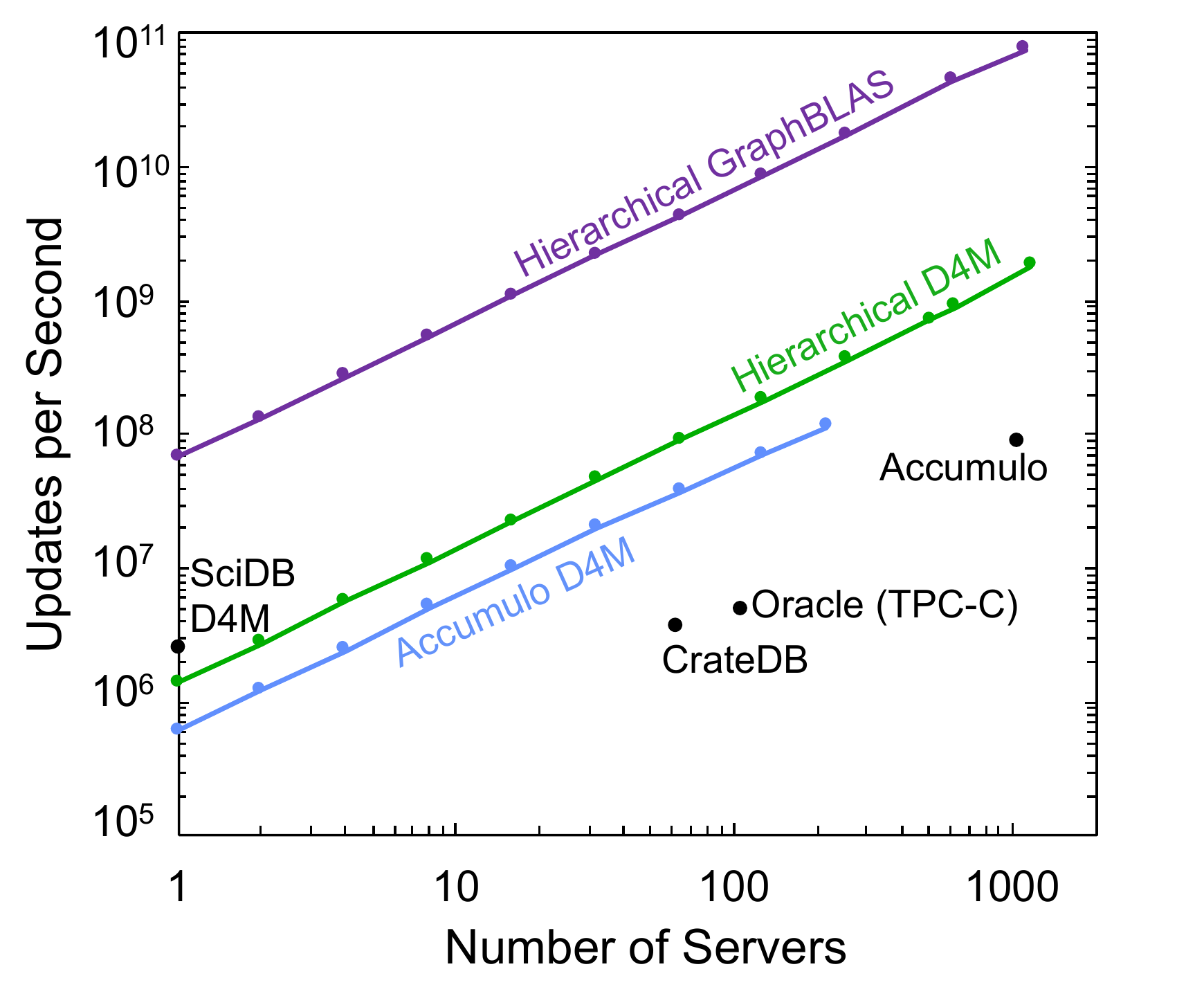}
\caption{Update rate as a function of number of servers for hierarchical GrapBLAS hypersparse matrices and other previous published work: Hierarchical D4M \cite{8547629}, Accumulo D4M \cite{kepner2014achieving}, SciDB D4M \cite{samsi2016benchmarking}, Accumulo \cite{sen2013benchmarking}, Oracle TPC-C benchmark, and CrateDB \cite{CrateDB}}
\label{fig:UpdateRate}
\end{figure}

  The scalability of the hierarchical hypersparse matrices are tested using a power-law graph of 100,000,000 entries divided up into 1,000 sets of 100,000 entries.  These data were then simultaneously loaded and updated using a varying number of processes on varying number of nodes on the MIT SuperCloud up to 1,100 servers with 34,000 processors. This experiment mimics thousands of processors, each creating many different graphs of 100,000,000 edges each.  In a real analysis application, each process would also compute various network statistics on each of the streams as they are updated.   The update rate as a function of number of server nodes is shown on Fig.~\ref{fig:UpdateRate}.  The achieved update rate of 75,000,000,000 updates per second is significantly larger than the rate in prior published results. This capability allows the MIT SuperCloud to analyze extremely large streaming network data sets.

\section*{Acknowledgement}

The authors wish to acknowledge the following individuals for their contributions and support: Bob Bond, Alan Edelman, Laz Gordon, Charles Leiserson, Dave Martinez, Mimi McClure, Victor Roytburd, Michael Wright.


%
%



\bibliographystyle{ieeetr}

\bibliography{HypersparseIngest}
%

\end{document}